\documentstyle{article}

 \def\test#1{}     

\def\beq{\begin{equation}}
\def\eeq{\end{equation}}

\def\beqx{\begin{displaymath}}
\def\eeqx{\end{displaymath}}

\def\beql{\arraycolsep .5mm \begin{eqnarray}}
\def\eeql{\end{eqnarray}}
\def\zeile{\nonumber \\[2mm] }   

\def\labelx#1{\label{#1}\test
                    {\raisebox{-4ex}{\makebox[0pt][r]{\tt #1}}}}

\def\rf#1{(\ref{#1})}

\def\c#1{{\cal #1}}
\def\ft#1#2{{\textstyle{{#1}\over{#2}}}}
\def\del{\partial}

\def\R{{\sf \bf R}}                    
\def\a{\alpha}
\def\b{\beta}
\def\g{\gamma}
\def\e{\epsilon}
\def\eps{\varepsilon}
\def\o{\omega}

\def\p{\psi}
\def\P{\Psi}

\def\t{{\theta}}

\def\grp#1({{\rm  #1}(}      
\def\alg#1({{\bf  #1}(}      

\def\M{{\c M}}    
\def\genus{n}     
\def\d{{\rm d}}   
\def\x{{\bf x}}   
\def\y{{\bf y}}
\def\deltaxy{{\delta (\x , \y )}}

\def\LG{\Gamma}      
\def\Pexp{{\c P} \, \exp\, }   

\def\euo#1#2{   e_{#1} ^{\;#2}  }

\def\euu#1#2{   e_{#1#2}  }
\def\Ao#1#2{   A_{#1} ^{\;#2}  }
\def\Au#1#2{   A_{#1#2}  }

\title{ Physical States in $d=3, N=2$ Supergravity}

\author{  B. de Wit \\\
          Institute for Theoretical Physics, Utrecht University,\\
          Princetonplein 5, 3508 TA Utrecht, The Netherlands \\\\
          H.-J. Matschull and H. Nicolai   \\\
          II. Institute for Theoretical Physics,
            University of Hamburg \\
          Luruper Chaussee 149, 22761 Hamburg, F.R.G.  }

\begin{document}
\maketitle

\begin{abstract}
To clarify some issues raised by D'Eath's recent proposal for
the physical states of $N=1$ supergravity in four dimensions,
we study pure (topological) $N=2$ supergravity in three
dimensions, which is formally very similar, but much easier to solve.
The wave functionals solving the quantum constraints can be
understood in terms of arbitrary functions on the space of
moduli and supermoduli, which is not Hausdorff. We discuss the
implications for the wave functionals and show that these
are not amenable to expansions in fermionic coordinates,
but can serve as lowest-order solutions to the quantum constraints
in an  expansion in $\hbar$ in more realistic theories.
\end{abstract}

Motivated by recent work on the nature of the full non-perturbative
wave functional of $d=4,N=1$ supergravity \cite{eath:93, page:93}
we study pure $d=3, N=2$ supergravity in this letter.
In contrast to supergravity in four dimensions, this theory
possesses only topological degrees of freedom,
so the problems associated with propagating degrees of freedom,
such as operator product singularities
and ordering ambiguities, which plague the canonical formulation
of supergravity in four dimensions, are absent. Moreover,
this model can be solved exactly at the quantum level,
just like $3d$ gravity \cite{Witten, Asht et al}. For these reasons,
and because the status of perturbative renormalizability versus
non-renormalizability of the $d=4,N=1$ is theory still pending, it
appears to be well suited to settle some of the issues raised in
\cite{eath:93}.

The reduced wave functionals of $N=2$ supergravity are given by
arbitrary functions depending on the moduli and supermoduli of
flat $SL(2,{\bf R})$ connections on a connected spatial surface
of arbitrary genus. However, the unreduced wave functionals
still depend on an infinite number of coordinates in the full
configuration space. From their structure it is immediately clear
that they are not amenable to the kind of expansion in terms
of the fermionic coordinates on which the arguments of
\cite{eath:93} are based. Furthermore, wave
functionals of this type also arise in four dimensions in the
limit $\hbar\to0$, where they can be used as input for a consistent
perturbative approach.
We note that pure $d=3, N=1$ supergravity was already discussed
in \cite{MN}. Here we prefer to consider
$N=2$ supergravity instead, mainly because the two Majorana
gravitinos can be combined into one Dirac vector spinor
and therefore admit a simpler representation of the fermionic
quantum operators. As a consequence, the associated quantum
constraints are formally very similar to those of $d=4,N=1$
supergravity as written down in \cite{eath:93}, apart from the
fact that we work in the so-called connection representation,
whereas \cite{eath:93} is based on the more familiar metric
formulation. Moreover the $N=2$ theory arises naturally in the
reduction of $d=4,N=1$ supergravity to three space-time
dimensions, which makes it an obvious starting point for this
study.

We use first-order formalism for the connection field, so
the basic fields are the dreibein $\euo \mu a$, the spin connection
$\Ao \mu a$, and the two-component complex gravitino field
$\psi_\mu$, which transforms as a spinor under $SL(2,\R)$,
corresponding to the spinor representation of the Lorentz group
$SO(1,2)$. The Lorentz-covariant derivatives and curvatures are
\beql
  D_\mu \euo \nu a &=& \del_\mu \euo \nu a - \eps^{abc}
                      \Au \mu b \,\euu \nu c \,,\zeile
  D_\mu \psi_\nu & = & \del_\mu \psi_\nu + \ft12 \Ao \mu a \g_a
                                        \psi_\nu  \,, \zeile
  F_{\mu\nu a} & = & \del_\mu \Au \nu a
                   - \del_\nu \Au \mu a
                   - \eps_{abc}\, \Ao \mu b \Ao \nu c\,.
\eeql
With this notation the Lagrangian of $N=2$ supergravity is given by
\beq
  \c L   =  \ft14 \eps^{\mu\nu\rho} \euo \mu a F_{\nu\rho a}
           + \eps^{\mu\nu\rho} \, \bar\psi_\mu D_\nu \psi_\rho\,.
\eeq
It is invariant under local Lorentz transformations
\beq
\delta \Ao \mu a = D_\mu \o^a \;\; , \;\;
\delta \p_\mu = -\ft12 \o^a \g_a \p_\mu\;\;,\;\; \delta \euo \mu
a = \varepsilon^{abc} \omega_b \,e_{\mu c}\,,
\eeq
and local supersymmetry transformations with parameter $\e$
\beq
\delta \Ao \mu a = 0 \;\; , \;\; \delta \p_\mu = D_\mu \e \;\;,
\;\; \delta \euo \mu a = \bar\e\g^a\p_\mu -\bar\p_\mu\g^a\e \,.
\eeq

Our conventions are as in \cite{MN, MN:93}. The
space-time manifold is assumed to be a direct product of a
connected two-dimensional manifold of genus $\genus$ (with local
coordinates $\x$) and the real line.
Greek indices $\mu, \nu,\dots$ denote coordinates
on the space-time manifold; with respect to the direct product
structure they take the values $t$ for the time coordinate and
$i,j, \dots$ for the local coordinates on the spatial manifold $\M$.
Flat $\grp SO(1,2)$ indices are designated by $a,b,\dots =0,1,2$.

The canonical treatment of the Lagrangian (1), which is explained
at length in \cite{MN, MN:93} \footnote{Standard
references on the canonical formulation of gravity are
\cite{ADM:62}.}, shows that the time components
$\euo ta$, $\Ao ta$, $\psi_t$ and $\bar\psi_t$
become Lagrange multipliers. They generate the constraints to be
given below. The components with spatial indices $i,j,\dots$ span
the phase space. After quantization one obtains the
(anti)commutation relations
\beql
   \big[ \Ao ia(\x) , \euo jb(\y)\big]
                &=& -2i\hbar\, \eps_{ij}\, \eta^{ab}\, \deltaxy
\,, \zeile
  \big\{  \psi_{i\alpha}(\x) , \bar \psi_{j\beta}(\y) \big\}
                &=&-i \hbar \,\eps_{ij}\, \delta_{\alpha\beta}\,
\deltaxy\,.
\labelx{dirac}
\eeql
(We will frequently suppress the $SL(2,\R)$ spinor
indices $\a,\b\dots$ on $\p_i$ and $\bar\p_i$.) An operator
realization
is easily found (this would be slightly more tricky for the
$N=1$ theory where the gravitino is Majorana). As the basic
variables, we take the connection $\Ao ia(\x)$ and $\p_i(\x)$.
Consequently, the wave functionals
are of the form $\P = \P [\Ao ia , \p_i ]$.
The canonically conjugate fields $\euo ia$ and $\bar \p_i$ are
represented by the functional differential operators
\beq
 \euo ia =  -2i\hbar\,\eps_{ij}\,  \frac \delta
                                 {\delta  \Au ja },
 \qquad
 \bar \psi_{i\alpha} =
           i \hbar \,\eps_{ij} \,\frac \delta{\delta \psi_{j\alpha} }.
\labelx{operators}
\eeq
Inserting these into the classical constraints we obtain the
quantum constraints
\beql
\left( i D_i (A) {{\delta}\over {\delta \Ao ia}} +
  \ft12 {{\delta}\over {\delta \p_i}} \g_a \p_i \right)\, \P &=&
0\,,
\labelx{Q-Lorentz} \\
\eps^{ij} F_{ija} (A) \,\P &=& 0 \,,
\labelx{Q-WDW} \\
\smallskip
\eps^{ij} D_i (A) \p_j \,\P &=& 0   \,,
\labelx{Q-S} \\
D_i (A) \frac \delta {\delta \p_i} \,\P &=& 0    \,.
\labelx{Q-S bar}
\eeql
Observe that there are no operator ordering ambiguities or
singularities in these expressions.
Solving quantum supergravity in three dimensions amounts
to solving these four functional differential equations.
The Lorentz constraint
\rf{Q-Lorentz} implies that $\P$ is invariant under local Lorentz
transformations and is thus trivially solved. The constraints \rf{Q-WDW}
and \rf{Q-S} tell us that $\P[A,\p ]$ has support on flat
$SL(2,\R)$ connections $\Ao ia$, and on gravitino fields $\p_i$
whose Rarita-Schwinger field strength vanishes. The former
constraint is just the Wheeler-DeWitt equation, which is implied
by the constraints \rf{Q-S} and \rf{Q-S bar}. This last
constraint requires $\P$ to be invariant under the
supersymmetry transformations $\delta \p_i =D_i \e$.

Unlike the constraints given in \cite{eath:93} for
supergravity in four dimensions, our constraints do not depend on
$\hbar$, which reflects the topological nature of the theory.
In both cases the supersymmetry constraints are of
first order and homogeneous in the fermionic operators.
Consequently, the fermionic constraints can be studied
separately on wave functionals $\P$ with definite fermion number.
Application of the constraints to the zero-fermion sector of
supergravity in four dimensions has lead to conflicting
conclusions \cite{eath:93,page:93}.

We now discuss the solutions of the quantum constraints,
following the analysis given in \cite{MN}.
We start with \rf{Q-WDW} and \rf{Q-S}, since
this is where most of the subtleties reside. For this
purpose, we find it convenient to
employ differential forms $A$ and $\p$ on $\M$ defined by
\beq
   A := \ft 12 \Ao ia \g_a \,\d x^i , \ \ \ \
   \psi := \psi_i \,\d x^i.
\eeq
Consequently, we must solve the conditions $F(A)=0$ and
$D(A)\p:=(d+A) \p=0$.
Clearly, these equations tell us that $A$ and $\p$ are pure gauge
locally, hence the absence of propagating degrees of freedom.
Globally, this need not be true in general, and the leftover
degrees of freedom are called ``moduli" (for $A$) and
``supermoduli" (for $\p$).
Locally $A$ and $\p$ are thus expressed by
\beq
A= g^{-1}\d g \,, \qquad \p = g^{-1}\d \phi \,,
\labelx{prepotential}
\eeq
where $g$ is an element of $SL(2,\R)$ and $\phi$ a fermionic
spinor function. Unlike $A$ and $\p$, $g$ and $\phi$ are not
necessarily globally defined (they are single-valued on the
covering manifold, however). Moreover they are only defined up to
multiplication of $g$ by a constant group element $h_0$ and shifts of
$\phi$ by a constant spinor $\e_0$. Hence $g,\phi$ and
$g',\phi'$ related by
\beq
g'(\x) = h_0^{-1} g(\x)  \;\;,\;\; \phi' (\x)= h_0^{-1} \big(
\phi (\x) - \e_0 \big)\,, \labelx{equivalence}
\eeq
are equivalent. We note also that under local Lorentz and
supersymmetry transformations $g$ and $\phi$ transform as
\beql
g(\x) &\longrightarrow & g(\x)\, h(\x)\,,  \zeile
\phi (\x) &\longrightarrow & \phi (\x) +   g(\x) \e (\x) \,,
\eeql
where $h(\x)$ and $\e(\x)$ are single-valued on $\M$.

A representation of $g$ and $\phi$ can be constructed as follows. Pick
an arbitrary point $\x_0 \in \M$ and let $\LG$ denote the first
fundamental group of $\M$ with base point $\x_0$. Given arbitrary
field configurations $A$ and $\p$ on $\M$ with vanishing
field strengths $F(A)$ and $D(A) \p$, we define $g(\x)$ and $\phi
(\x)$ by
\beq
g(\x) := \Pexp \int_{\x_0}^{\x} A \;\; , \;\;
\phi (\x ) := \int_{\x_0}^{\x} g \p
\labelx{g,phi}
\eeq
These expressions depend on the base point $\x_0$, but are
insensitive to continuous deformations of the path connecting
$\x_0$ to $\x$.
Therefore they are affected by local Lorentz and
supersymmetry tranformations at the base point, which induce the
transformations \rf{equivalence} with $h_0 \equiv h(\x_0 )$ and
$\e_0 \equiv \e (\x_0 )$. Likewise, changing the basepoint and
thus the path connecting it to $\x$, changes $g$ and $ \phi$
in accord with \rf{equivalence}.

For $\g \in \LG$ and $f(\x)$ an
arbitrary and not necessarily single-valued function on $\M$, we
denote by $f(\x_0 + \g)$ the value of $f$ obtained by starting at
$\x_0$ and letting $\x$ traverse the loop $\g$ once. Assuming
single-valuedness for $A$ and $\p$ implies that the effect of
traversing the loop may lead to a deficit of the form
\rf{equivalence}. The bosonic and fermionic holonomies $g_\g$ and
$\phi_\g$ parametrize this deficit and are thus defined by
\beq
g(\x_0 + \g) = g_\g \,g(\x_0) \;\; , \;\;
\phi (\x_0 + \g ) = \phi_\g  + g_\g \,\phi (\x_0 )
\labelx{holonomies}
\eeq
Under local Lorentz and supersymmetry transformations $g_\g$ and
$\phi_\g$ are invariant. As $g$ and $\phi$ are only defined up
to the transformations \rf{equivalence}, the bosonic and
fermionic holonomies related by
\beq
g_\g' = h_0^{-1} g_\g h_0  \;\; , \;\;
\phi_\g' = h_0^{-1} \big( \phi_\g + (g_\g -1) \e_0 \big)
\labelx{holotrafo}
\eeq
should be identified.
Therefore the wave functionals depend only on the conjugacy
classes with respect to \rf{holotrafo}. (This can for instance
be ensured by choosing a reduced wave functional that is
invariant with respect to the tranformations \rf{holotrafo}.)

The space of moduli and supermoduli on a spatial (Riemann) surface
of genus $n$ is now defined as the space of
holonomies \rf{holonomies} modulo the transformations
\rf{holotrafo} and the constraint
\beq
\prod_{j=1}^n \a_j \b_j \a^{-1}_j \b^{-1}_j = 1
\labelx{abab}
\eeq
defining the first fundamental group on a Riemann surface of
genus $n$, where $\a_j , \b_j$ ($j=1,\dots,n$) constitute the usual
basis of homology cycles on the Riemann surface \cite{FK}. More
rigorously, the space of (super)moduli consists of the set of
conjugacy classes of group homomorphisms from the fundamental
group into the gauge group (the combined group of local Lorentz
and supersymmetry transformations). This homomorphism is defined
by \rf{g,phi}.
Condition \rf{abab} thus translates into similar constraints
on the corresponding quantities $(g_{\a_j},\phi_{\a_j})$ and
$(g_{\b_j},\phi_{\b_j})$, because
$g(\x)$ and $\phi(\x)$ must be single-valued around the
contractible loop defined by \rf{abab}. Here one must use the
composition law induced by the homomorphism,
\beq
g_{\g_1\circ\g_2} = g_{\g_1}\,g_{\g_2} \,,\qquad
\phi_{\g_1\circ\g_2}= \phi_{\g_2} + g_{\g_2}\,\phi_{\g_1} \,,
\labelx{composition}
\eeq
where we used that $\phi$ vanishes at the base point.

Detailed discussions of the bosonic moduli space of flat
$SL(2,\R)$ connections may be found in \cite{Asht et al} and
and \cite{Verlinde}. However, the first reference discusses only
elliptic conjugacy classes of $SL(2,\R )$, while the second
deals only with hyperbolic conjugacy classes, which are shown
to be directly related to Teichm\"uller space\footnote{It would
be interesting to see whether the space of supermoduli as defined
above can be similarly related to the super-Teichm\"uller space
of a Riemann surface if one restricts the bosonic conjugacy
classes to the hyperbolic sector. The dimensions of these spaces
are the same.}. On the other hand, it could be plausibly argued that
any discussion of $3d$ quantum supergravity should take into account
{\it all} of (super)moduli space (see also \cite{marolf}). An
unexpected property of the bosonic moduli space is that it is not
a Hausdorff space in general \cite{ashtekar:93}. This feature is
usually related to the fact
that moduli space is defined as the quotient of two
{\em infinite} dimensional spaces, and has been known to
mathematicians for a long time \cite{Newstead}; we
will explain it in terms of an elementary example at the
end of this paper when discussing the torus.
It also has implications for the fermionic moduli and for
the wave functionals. In particular, the presence of
extra fermionic moduli seems to be correlated with the lack
of the Hausdorff property. To explain this point, let us count the
number of fermionic moduli.
There are $2n$ $\phi_\g$'s, each with two spinor components,
hence altogether $4n$ fermionic holonomies. Since $\phi (\x)$ must
be single-valued when transported around the contractible
loop \rf{abab}, only $4n-2$ of them are independent (see the
discussion above). Moreover, for generic bosonic holonomies, we can
use \rf{holotrafo} to gauge away two more fermionic holonomies,
so in general there will be $4n-4$ fermionic moduli. For
non-generic $g_\g \in SL(2,\R)$, the matrices $(g_\g -1)$
may not be invertible. From \rf{holotrafo}, it is evident that
we can still remove two fermionic degrees of freedom
as long as there remains at least one homology cycle $\g$, for which
$(g_\g -1)$ is invertible. Otherwise, we cannot use
\rf{holotrafo} to gauge away fermionic holonomies, and at such
non-generic points, there will be more fermionic moduli. This means
that the ``superspace" spanned by the bosonic and fermionic moduli
is not a direct product space, but rather more like a sheaf!
The existence of such singular points is a feature which is entirely
due to the non-compactness of $SL(2,\R)$, since invertibility
may only fail for parabolic conjugacy classes. Since all matrices
$(g_\g -1)$ must be non-invertible for extra fermionic moduli
to exist, the singular points form a set of very low
dimension and therefore become ``less and less important" with
increasing genus $n$. Nevertheless, we still face the
obvious question how to define wave functionals on such a space.
If we insist on continuity, these must be constant along those
bosonic moduli for which the Hausdorff property breaks down;
this is also the point of view adopted in \cite{ashtekar:93}.
For the same reason, they could not depend then on the extra
fermionic moduli, either. One can also avoid the singular points
altogether by restricting the space of bosonic moduli to the
hyperbolic sector from the outset as proposed in \cite{Verlinde}.
In any case, different prescriptions can be expected to lead to
inequivalent theories of quantum (super)gravity.

The physics content of the theory is completely encapsulated in
the wave functions $f(g_{\a_j}, g_{\b_j}, \phi_{\a_j}, \phi_{\b_j})$
depending on the conjugacy
classes of the fermionic and bosonic holonomies as defined by
\rf{holotrafo} and the constraint \rf{abab}.
Since these moduli and supermoduli form a finite dimensional
space at each genus, all further manipulations are in principle
well-defined. So, we can now define a scalar product by means of
a suitable measure and evaluate the observables introduced in
\cite{MN} on the states. We note, however, that it is a priori not
clear what measure to choose due to the non-compactness of the group
$SL(2,\R)$ and how to define the class of admissible functions
$f$ (see e.g. \cite{marolf}).

We can equivalently describe the solutions in terms of wave functionals
on the full configuration space spanned by $\Ao ia (\x)$ and
$\p_i (\x)$. The resulting expression is somewhat cumbersome and
reads
\beql
\P [\Ao ia, \p_i ] &=& \int^\prime  \prod_{j=1}^n \Big\{ \d g_{\a_j}\,
\d\phi_{\a_j} \delta \Big(g_{\a_j} -\Pexp \oint_{\a_j} A \Big)
 \delta \Big(\phi_{\a_j} - \oint_{\a_j} g\,\p \Big) \Big\} \zeile
&& \,\times \prod_{j'=1}^n\Big\{\d g_{\b_{j'}}\,  \d\phi_{\b_{j'}}
\delta \Big(g_{\b_{j'}} -\Pexp \oint_{\b_{j'}} A \Big) \delta
\Big(\phi_{\b_{j'}} - \oint_{\b_{j'}} g\,\p \Big) \Big\} \zeile
&&\,\times f\big(g_{\a_j},g_{\b_j} , \phi_{\a_j}, \phi_{\b_j}
\big) \; \prod_{\x} \,\delta \big(F(A(\x))\big)\;\;
\delta\big(\varepsilon^{ij}\,D_i(A)\p_j (\x)\big)  \,,
\labelx{3dwavefunctional}
 \eeql
where $\d g$ denotes the Haar measure on $SL(2,\R)$ and $\d \phi$
are Grassmann integrals; the prime attached to the
(finite-dimensional) integral is to indicate that it is to be
performed only over those holonomies satisfying the constraints
following from \rf{abab}. The delta functionals in the last line
enforce the quantum constraints \rf{Q-WDW} and \rf{Q-S}. The fermionic
delta functional is defined as the infinite product over its
arguments in the usual way. Evidently, this makes an expansion
of the wave functional in terms of fermionic coordinates starting ``at
the bottom" or ``at the top" of the Dirac sea rather pointless,
because the sea is neither empty nor completely filled. Applying the
last constraint \rf{Q-S bar} to the above wave functional we get two
contributions. On the fermionic delta functional (last line of
\rf{3dwavefunctional}), it produces a curl $\varepsilon^{ij}
D_i D_j$, which vanishes by $F(A) =0$, while on the delta functions
(second line of \rf{3dwavefunctional}), it gives zero
because the moduli are locally supersymmetric. Similarly,
\rf{3dwavefunctional} is annihilated by \rf{Q-Lorentz} because
the delta functionals are, and the remaining part of the integral
involves only manifestly Lorentz invariant quantities. Thus, the
wave functional \rf{3dwavefunctional} satisfies all constraints.

Obviously, the above expression is
highly formal and at best of heuristic value; rigorous
manipulations can only be performed in terms  of the functions $f$.
With this in mind, we now return to the constraints of
$d=4,N=1$ supergravity as written down in \cite{eath:93}.
 They read
 \beql
&& \left(\varepsilon^{ijk}e_{AA'i}(\x)\, D_j\p^A_k(\x)  -\ft12
\hbar \kappa^2  \p^A_i(\x)\,  {\delta\over {\delta
e^{AA'}_i(\x)}} \right) \P [e,\p ]  =0  \,,   \zeile
&& \left(D_i {\delta\over \delta \p^A_i(\x)} +\ft12 \hbar\kappa^2
{\delta\over \delta e^{AA'}_i(\x)}\, D^{BA'}_{ji}\, {\delta\over
\delta\p^B_i(\x)}\right)\P[e,\p ]  =0 \,,  \labelx{death}
 \eeql
where the notation has been appropriately changed to four dimensions
(with two-component spinor notation).
It is not relevant for our arguments that \cite{eath:93} uses the
dreibein (or metric) representation rather than the connection
representation that we have been using up to this point. Observe
that these constraints depend on $\hbar$, which reflects the
presence of propagating degrees of freedom in this theory. Obviously,
these constraints are much harder to solve, because both the
dreibein and its conjugate momentum appear in \rf{death}, so the
constraint operator is not just a multiplication operator as
in the $3d$ theory. The dependence of \rf{death} on $\hbar$
suggests an expansion of $\P$ in terms of $\hbar$. In analogy
with \rf{3dwavefunctional} the first of the constraints
\rf{death} is schematically solved by
\beq
 \P [e,\p ] = \prod_{\x}\; \delta
\Big(R^{(3)}\big(e(\x)\big)\Big)\; \delta \Big(
 D\big(\omega(e(\x)\big)\p (\x) \Big)
          \; F({\rm ``moduli"}) + O(\hbar ) \labelx{4dwavefunctional}
\eeq
to lowest order in $\hbar$, where by ``moduli" we mean all the
bosonic and fermionic degrees of freedom contained in the
dreibein and gravitino which are not killed by the first constraint
(there are now infinitely many such moduli, because we have propagating
modes). Note that the argument of the delta functionals are just the
$O(\hbar^0 )$ parts of the Wheeler-DeWitt operator and the
first supersymmetry constraint in \rf{death}, respectively. To lowest
order, the second constraint then annihilates the wave functional
\rf{4dwavefunctional} as well, because application of the
$O(\hbar^0 )$ term produces a commutator of two covariant
derivatives, which after a little calculation is found to
be proportional to the curvature scalar, and thus
vanishes on \rf{4dwavefunctional} \footnote{For this argument
we have to make use of the formula
\beq
R_{ijkl} = g_{ik} R_{jl} - g_{jk} R_{il} - g_{il} R_{jk}
       + g_{jl} R_{ik} + \ft12 R \big( g_{ik} g_{jl} -
            g_{jk} g_{il} \big) ,
\eeq
which is valid in three dimensions only.}.
Apart from the highly singular nature of the $O(\hbar^0)$ term,
we again see that an expansion in the fermionic coordinates
does not seem feasible. Just as in
\rf{3dwavefunctional}, the fermionic occupation is such that
$\P$ starts ``in the middle" of the Dirac sea at infinite distance
from both its top and bottom, but there will now be further terms
at higher orders in $\hbar$. With \rf{4dwavefunctional} as
lowest-order solution we can thus apply perturbation theory in
$\hbar$, at least in principle. In practice, matters become quite
complicated and we have not attempted to carry these
considerations any further. We suspect that the
short-distance singularities, which were so conspicuously absent
from \cite{eath:93}, will reappear in higher orders.

We conclude this paper by briefly discussing the space of moduli
and supermoduli for the torus and explicitly demonstrating that
this space is not Hausdorff.  There are two non-trivial
homology cycles $\a$ and $\b$. The condition \rf{abab} and the
corresponding condition for the fermionic holonomies (cf.
\rf{composition}) reduce to
\beq
g_\a g_\b g_\a^{-1} g_\b^{-1} =1  \;\;\; , \;\;\;
(g_\a -1) \phi_\b = (g_\b -1) \phi_\a
\labelx{toruscond}
\eeq
Two sets of holonomies are then equivalent if they are related as
in \rf{holotrafo}. From the first condition in \rf{toruscond}, we
infer that $g_\a$ and $g_\b$ must commute, and hence belong to the same
conjugacy class of $SL(2,\R)$. There are four special points
corresponding to the matrices $g_\a , g_\b = \pm
{\bf 1}$. Only for $g_\a=g_\b=\bf1$ there are fermionic moduli.
In principle there are four such moduli (corresponding to the
four constant $\p_i$ coordinates), but by requiring equivalence
with respect to $SL(2,{\bf R})$ we are left with only two. Apart from
these four points there are three types of conjugacy classes
which we now discuss in turn.
\begin{itemize}
\item
If both $g_\a$ and $g_\b$ are elliptic (corresponding to mutually
commuting timelike generators), they can always be brought into
form of an $SO(2)$ transformation,
\beq
    g_\g = \pmatrix{ \cos \t_\g & \sin \t_\g \cr
                \noalign{\vskip 2mm}
                   -\sin \t_\g & \cos \t_\g \cr} \ \ \ , \ \ \
    \phi_\g = 0
\labelx{elliptic}
\eeq
for $\g = \a , \b$.
As $(g_\g -1)$ is always invertible in this case, it is easy to
see that all fermionic holonomies can be gauged away and there
are thus no supermoduli. From the
representation \rf{elliptic} it is evident that this part of
the bosonic moduli space is just a torus with the four points
corresponding to $g_\a, g_\b = \pm {\bf 1}$ cut out.
\item
If both $g_\a$ and $g_\b$ are hyperbolic (corresponding to
mutually commuting spacelike
generators), the standard representative is given by
\beql
 g_\a &=& \pm \pmatrix{ e^{r\cos \ft\t  2 } & 0 \cr
                     \noalign{\vskip 2mm}
                        0 & e^{-r \cos \ft\t  2 } \cr}, \ \ \
 g_\b = \pm \pmatrix{ e^{r\sin \ft\t  2 } & 0 \cr
                       \noalign{\vskip 2mm}
                      0 & e^{-r \sin \ft \t  2 } \cr}, \zeile
  \phi_\g &=&0,
\labelx{hyperbolic}
\eeql
for $r > 0$ and $0\leq \t < 2\pi$. The absence of fermionic moduli
is again straightforward to prove. This part of moduli space
consists of four copies of the plane with the origin cut out.
\item
If both $g_\a$ and $g_\b$ are parabolic (corresponding to
mutually commuting lightlike
generators), we get
\beq
 g_\a = \pm\pmatrix{ 1 & a  \cr 0 & 1 \cr} , \ \ \ \
 g_\b = \pm\pmatrix{ 1 & b  \cr 0 & 1 \cr},
\labelx{parabolic}
\eeq
where we can scale $a$ and $b$ by an $SL(2,{\bf R})$ transformation
to obey the restriction $a^2 + b^2 =1$. This part of the bosonic
moduli space is thus homeomorphic to four copies of the circle
$S^1$ (with no points cut out). In contrast to the cases
discussed before, however, the matrices $(g_\g -1)$ fail to be
invertible if both signs are +, and we then have extra fermionic
moduli.
\end{itemize}

To see that this is not a Hausdorff space consider for instance
a sequence of elliptic elements in $SL(2,\R )$ approaching a
point on the lightlike boundary (i.e., on $S^1$
according to the above analysis). By conjugating these
matrices into the $SO(2)$ subgroup as we did above, we obtain a
sequence of points in $SO(2)$ converging to $\pm {\bf 1}$. However,
although they converge to different limit points on the lightlike
boundary, these two sequences are identified in the moduli space.
Phrased otherwise, this means that for any
two distinct points in the parabolic sector, we cannot find open
neighbourhoods that separate them! The situation is similar for
sequences approaching the boundary in the hyperbolic sector,
although there are now {\em three} limit points, two on $S^1$ (at
opposite points) and $\pm\bf 1$. Thus, the total moduli space
can be modelled as follows: take a torus with four holes and four copies
of the plane with a hole cut out at the origin. Glue them together
in such a way that each plane gets attached to one of the holes in
the torus and that each boundary on the torus winds twice around the border
of the hole in the plane to which it is attached. These
boundaries have the topology of a circle and parametrize the
four parabolic conjugacy classes. To each of them we put an extra
point and define the open neighborhoods of these points
to be all open sets containing the whole circle. There are no
fermionic moduli except along one of the circles.

Finally, we note that the difficulties encountered here disappear
altogether in the so-called mini-superspace approximation, where
from the outset one deals only with a finite number of
degrees of freedom \cite{mini}. In the light of the results obtained
it appears that this approximation cannot truly capture the
remarkable features of quantum gravity and supergravity. \\
\noindent
{\it Note added:} The inapplicability of an expansion in terms of
``fermion number" has also been demonstrated in \cite{Freedman},
which is based on an analysis of the physical state functionals
of $d=4,N=1$ supergravity in terms of the free spin-3/2 field.

\vspace{6 mm}

\noindent
{\bf Acknowledgements:} We benefitted from valuable discussions with
D.Z.~Freedman, D.N.~Page and H.~Verlinde. We would also like to
thank P. Slodowy for informing us about reference \cite{Newstead}.


\begin{thebibliography}{1}

\bibitem{eath:93}
P.D. D'Eath.
\newblock {\it Physical states in $N=1$ supergravity},
\newblock Preprint DAMTP-R-93-4, University of Cambridge, 1993.

\bibitem{page:93}
D.N. Page.
\newblock {\it Inconsistency of D'Eath's bosonic states of $N=1$
supergravity},
\newblock Preprint~Thy-28-93, University of Alberta, 1993.

\bibitem{Witten}
E. Witten, Commun. Math. Phys. {\bf 121} (1989) 351.

\bibitem{Asht et al}
A. Ashtekar, V. Husain, C. Rovelli, J. Samuel, L. Smolin,
Class. Quantum Grav. {\bf 6} (1989) L185.

\bibitem{MN}
H. Nicolai and H.J. Matschull, J. Geom. Phys. {\bf 11} (1993) 15

\bibitem{MN:93}
H.J. Matschull and H. Nicolai, preprint DESY 93-073, gr-qc/9306018,
         to appear in Nucl. Phys. B.

\bibitem{ADM:62}
R. Arnowitt, S. Deser, and C.W. Misner, in {\it Gravitation: An
Introduction to Current Research}, ed. L. Witten,
  Wiley, New York, 1962; \\
C.W. Misner, K.S. Thorn, and J.A. Wheeler, {\it Gravitation},
Freeman, New York, 1973.

\bibitem{FK}
H. Farkas and I. Kra, {\it Riemann Surfaces}, Graduate Texts in
    Mathematics 71, Springer Verlag, Berlin, 1992

\bibitem{Verlinde}
H. Verlinde, Nucl.Phys. {\bf B337} (1990) 652.

\bibitem{marolf} D.M. Marolf, {\em An Illustration of $(2+1)$
Gravity Loop Transform Troubles}, Syracuse preprint SU-GP-93-3-4A,
1993.

\bibitem{ashtekar:93} A. Ashtekar and J. Lewandowski, Class.
Quantum Grav. {\bf 10} (1993) L69.

\bibitem{Newstead}
P. Newstead, {\it Introduction to Moduli Problems and Orbit Spaces},
Tata Institute Lecture Notes Vol. 51, Springer Verlag, Berlin, 1978.

\bibitem{mini}
S. Elitzur, A. Forge and E. Rabinovici,
       Nucl.Phys. {\bf B274} (1986) 60; \\
P.D. D'Eath and D.I. Hughes, Phys. Lett. {\bf 214B} (1988) 498; \\
R. Graham, Phys. Rev. Lett. {\bf 67} (1991) 1381,
           Phys. Lett. {\bf B277} (1992) 393;     \\
P.D. D'Eath, S.W. Hawking and O. Obregon,
       Phys. Lett. {\bf B300} (1993) 44


\bibitem{Freedman}
S. Carroll, D.Z. Freedman, M. Ortiz and D.N. Page, {\it Physical
    States in Supergravity}, preprint in preparation

\end{thebibliography}
\end{document}